\documentclass[aps,prc,preprint,groupedaddress,showpacs]{revtex4}
\usepackage{graphicx}% Include figure files

\begin{document}
%\draft

\def\dj{\leavevmode\setbox0=\hbox{d}\dimen0=\wd0
        \setbox0=\hbox{d}\advance\dimen0 by -\wd0
        \rlap{d}\kern\dimen0\hbox to \wd0{\hss\accent'26}}
\def\DJ{\leavevmode\setbox0=\hbox{D}\dimen0=\wd0
        \setbox0=\hbox{D}\advance\dimen0 by -\wd0
        \rlap{D}\kern\dimen0\hbox to \wd0{\raise -0.4ex\hbox{\accent'26}\hss}}

\title{$^{4}$He decay of excited states in $^{14}$C}

\author{N. Soi\'{c}}
%\affiliation{
%School of Physics and Astronomy, University of Birmingham, Edgbaston, Birmingham~B15~2TT, U.~K.}
\altaffiliation{Permanent address: Ru\dj er Bo\v{s}kovi\'{c} Institute, Bijeni\v{c}ka 54, 
HR-10000 Zagreb, Croatia}
%\email[email:]{soic@lnr.irb.hr}

\author{M.~Freer}
%\affiliation{
%School of Physics and Astronomy, University of Birmingham, Edgbaston, Birmingham, B15~2TT, U.~K. }

\author{ L. Donadille}
\altaffiliation{Present address: CEA-Saclay, DAPNIA/SPhN, Bt. 703, Pice 162, F-91191 Gif sur Yvette
 Cedex, France}
%\affiliation{
%School of Physics and Astronomy, University of Birmingham, Edgbaston, Birmingham~B15~2TT, U.~K. }

\author{  N. M. Clarke}
%\affiliation{
%School of Physics and Astronomy, University of Birmingham, Edgbaston, Birmingham~B15~2TT, U.~K. }

\author{ P. J. Leask}

\affiliation{
School of Physics and Astronomy, University of Birmingham, Edgbaston, Birmingham, B15~2TT, U. K. }

\author{  W. N. Catford }
\author{  K. L. Jones}
\altaffiliation{Present address: GSI, Gesellschaft f\"{u}r Schwerionenforschung mbH, Planckstrasse 1,
 D-64291 Darmstadt, Germany}
\author{ D. Mahboub }
\affiliation{
School of Electronics and Physical Sciences, University of Surrey, Guildford, Surrey, GU2 5XH, U. K.}

\author{ B. R. Fulton }
\author{B. J. Greenhalgh}
\author{ D. L. Watson}
\affiliation{
Department of Physics, University of York, Heslington, York, YO10 5DD, U. K.}

\author{D. C. Weisser}
\affiliation{Department of Nuclear Physics,  The Australian National University, Canberra ACT 0200, Australia }

\date{ \today}

\begin{abstract}
A study of the $^7$Li($^{9}$Be,$^4$He$^{10}$Be)$^2$H reaction at $E_{beam}$=70 MeV
has been performed using resonant particle spectroscopy techniques and provides
the first measurements of $\alpha$-decaying states in $^{14}$C. Excited states
are observed at 14.7, 15.5, 16.4, 18.5, 19.8, 20.6, 21.4, 22.4 and 24.0 MeV.
The experimental technique was able to resolve decays to the various particle
bound states in $^{10}$Be, and provides evidence for the preferential decay of
the high energy excited states into states in $^{10}$Be at $\sim$6 MeV. The
decay processes are used to indicate the possible cluster structure of the
$^{14}$C excited states.
\end{abstract}


\maketitle

\section{Introduction}
The study of clustering within nuclei has received renewed interest in the
light of the potential impact that the underlying cluster structure has on
valence particles, particularly in neutron-rich nuclei. The nucleus $^{8}$Be
has a pronounced $\alpha$-$\alpha$ cluster structure, as evidenced by the
deformed rotational band built on the ground state and the large
$\alpha$-particle decay width \cite{Buc77,Hiu72}. Thus, the wave-functions of
neutrons introduced into this system are markedly affected by the two-centre
nature of the underlying potential. This situation is reminiscent of the atomic
binding of covalent molecules where electrons move in multi-centre orbits
characterised by the locations of the nuclei. For example, in the binding of
O$_{2}$ the valence electrons reside in orbits with $\sigma$ and
$\pi$-character, since such orbits result from the linear combination of
$p$-type orbits which the atomic electrons occupy. Similarly, valence
neutrons in the neutron-rich Be isotopes could occupy $\sigma$ and $\pi$-type
orbits given that when separated the neutrons would reside in orbits in the
helium nuclei with $p$-type characteristics.

Indeed, these ideas have been developed considerably in the case of the Be
isotopes and there appears to be reasonably good evidence for states with both
the $\sigma$ and $\pi$ characteristics in $^{9}$Be and $^{10}$Be
\cite{von96,von97,von97a}. There is also tentative evidence for the extension
of these ideas to $^{11}$Be \cite{von96,von97,von97a} and $^{12}$Be
\cite{Fre99,Fre01}. Somewhat remarkably these characteristics are also found in
the calculations using the Antisymmetrised Molecular Dynamics (AMD) framework,
a model which in principle contains no explicit cluster, or molecular, content
\cite{Dot97,Kan99,Kan02}.

An extension to such ideas is the formation of three-centre molecules,
principally formed from three $\alpha$-particles, i.e. $^{12}$C. However, the
building block for such linear molecules, the 3$\alpha$ chain state remains to be
experimentally observed. There has been a long association between the 7.65 MeV
(0$^+$) state and the chain configuration. However, it is now generally
accepted, due to the lack of experimental evidence for a rotational 2$^+$
state, that this state is probably linked with a triangular arrangement of the
three clusters \cite{Fed96}.  The next possible candidate for the chain is the
10.3 MeV state. However, this resonance is broad and lies in a region where the
spectroscopy of $^{12}$C is extremely complex. Hence, there is no convincing
evidence for the existence of the 3$\alpha$-chain. Indeed, molecular orbit
model (MO) calculations, which reproduce well the properties of $^{10}$Be
\cite{Ita00}, have found that the 3$\alpha$ state is unstable against the
bending mode \cite{Ita01}. The same calculations \cite{Ita01,Ita02} suggest
that the introduction of valence neutrons may stabilize the chain structure.

The present paper presents an experimental study of the
$^7$Li($^{9}$Be,$^{14}$C$^*$$\rightarrow^{10}$Be+$\alpha$)$^2$H reaction, in which
the $\alpha$-decays of $^{14}$C excited states both to the $^{10}$Be ground
state and excited states are observed. These $^{10}$Be states have previously
been characterised in terms of their molecular structure
\cite{von96,von97,von97a}.

\section{Experimental Details}

The measurements were performed at the Australian National University's 14UD
tandem accelerator facility. A 70 MeV $^9$Be beam, of intensity 3 enA, was
incident on a 100 $\mu$g cm$^{-2}$ Li$_2$O$_3$ foil. The integrated beam
exposure was 0.45 mC.

Reaction products formed in reactions of the $^9$Be beam with the target were
detected in an array of four charged particle telescopes. These telescopes
contained three elements which allowed the detection of a wide range of
particle types, from protons to $Z$=4 to 5 nuclei. The first elements were
thin, 70$\mu$m, 5$\times$5 cm$^{2}$ silicon detectors segmented into four
smaller squares (quadrants). The second elements were position-sensitive strip
detectors with the same active area as the quadrant detectors, but divided into
16 position-sensitive strips. These strips were arranged so that the position
axis gave maximum resolution in the measurement of scattering angles. Finally,
2.5 cm thick CsI detectors were used to stop highly penetrating light
particles. These detector telescopes provided charge and mass resolution up to
Be, allowing the final states of interest to be unambiguously identified. The
position and energy resolution of the telescopes was $\sim$1 mm and 300 keV,
respectively. Calibration of the detectors was performed using elastic
scattering measurements of $^{9}$Be from $^{197}$Au and $^{12}$C targets. The
four telescopes were arranged in a cross-like arrangement, separated by
azimuthal angles of 90 degrees. Two opposing detectors were located with their
centres at 17.3 and 17.8 degrees (detectors 1 and 2) from the beam axis and
with the strip detector 130 mm from the target, whilst the remaining pair were
at the slightly larger angles of 28.6 and 29.7 degrees (detectors 3 and 4), 136
mm from the target.

\section{Results}
\begin{figure}
\includegraphics[width=0.75\textwidth]{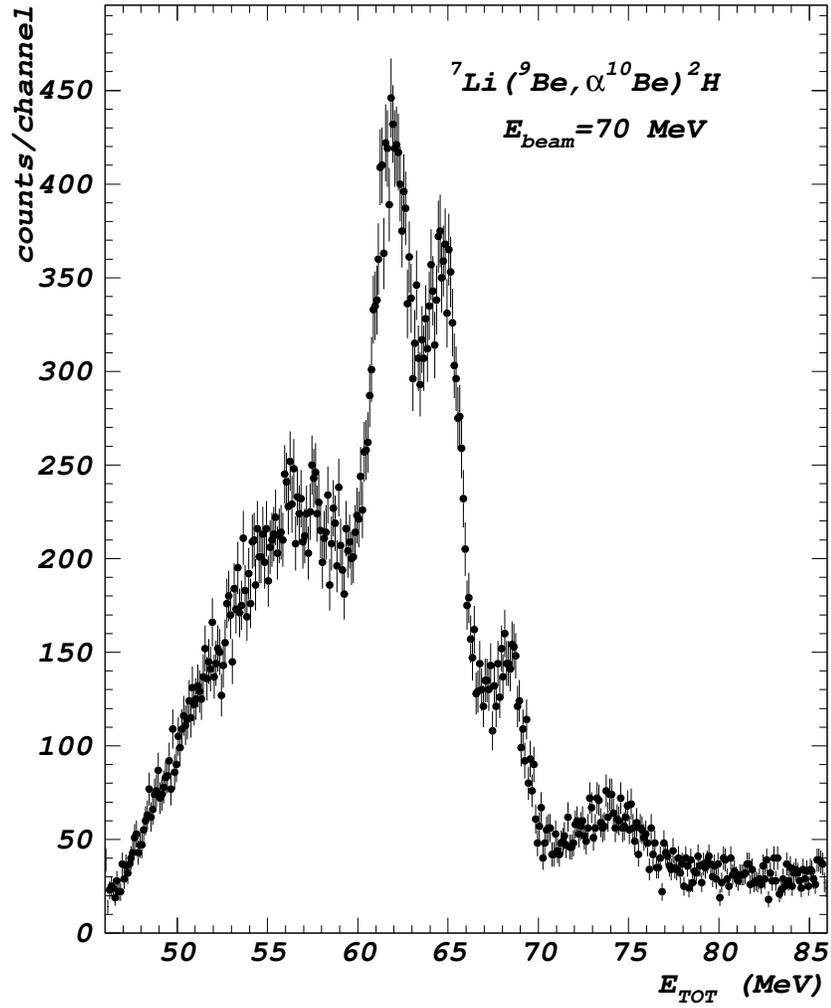}
\caption{\label{etot} Total energy spectrum for $^{4}$He+$^{10}$Be coincidences,
assuming a $^2$H recoil. Only statistical uncertainties are plotted.}
\end{figure}

The resolution of $^4$He and $^{10}$Be locii in the spectra of the particle
identification telescopes, and the measurement of the energies and angles of
these detected particles permitted the kinematics of the
$^7$Li($^{9}$Be,$^{10}$Be$\alpha$)$^2$H reaction (Q=-1.91 MeV) to be fully
reconstructed. Figure \ref{etot} shows the total energy spectrum
($E_{tot}=E_{Be}+E_{\alpha}+E_d$) for the two smaller angle detectors (1 and 2)
for one of the two possible coincidence combinations. There are three clear
peaks in this spectrum. The two highest energy peaks are separated by 3.4 MeV
and the third by a further 2.6 MeV. These would thus correspond to all of the
bound states in $^{10}$Be. The highest energy peak ($E_{tot}$ = 68.1 MeV)
corresponds to the production of $^{10}$Be in the ground state, the next
 lower ($E_{tot}$ = 64.7 MeV) is the 3.4 MeV, 2$^+$ excitation, and the lowest
 energy peak  ($E_{tot}$ $\approx$ 62 MeV) in this total energy spectrum
corresponds to the population of the 5.958 MeV (2$^+$), 5.960 MeV (1$^-$),
6.179 MeV (0$^+$) and 6.263 MeV(2$^-$) states, which are unresolved in the
present spectrum given the energy separation is only $\sim$300 keV. The energy
resolution in the total energy spectrum is $\sim$2.0 MeV, i.e. worse than the
intrinsic energy resolution of the telescopes. This inferior resolution is due
to the very light mass of the recoil particle, since a relative small
uncertainty in the measured momenta of the reaction products translates into a
large uncertainty in the energy of the unobserved deuteron. Nevertheless, the
resolution is sufficient to resolve the $^{10}$Be spectrum into three
components; the ground state, the 2$^+$ state and the group of states at
$\sim$6 MeV. The spectrum also shows a broad bump at higher total energies,
which has been identified as arising from $^{9}$Be+$\alpha$ coincidences
from the $^{7}$Li($^{9}$Be,$^{9}$Be$\alpha$)$^{3}$H reaction
leaking through the A=10 mass selection windows. This association has been
confirmed by the analysis of this latter reaction using the appropriate
 particle masses. There is also a broad bump at
lower energies which partially extends under the peaks. This contribution is
from more complex final states (see later). Reactions from the oxygen and the
inevitable carbon components in the targets have very negative Q-values ($<$-19
MeV) and are thus not represented in the present spectrum.

\begin{figure}
\includegraphics[width=0.65\textwidth]{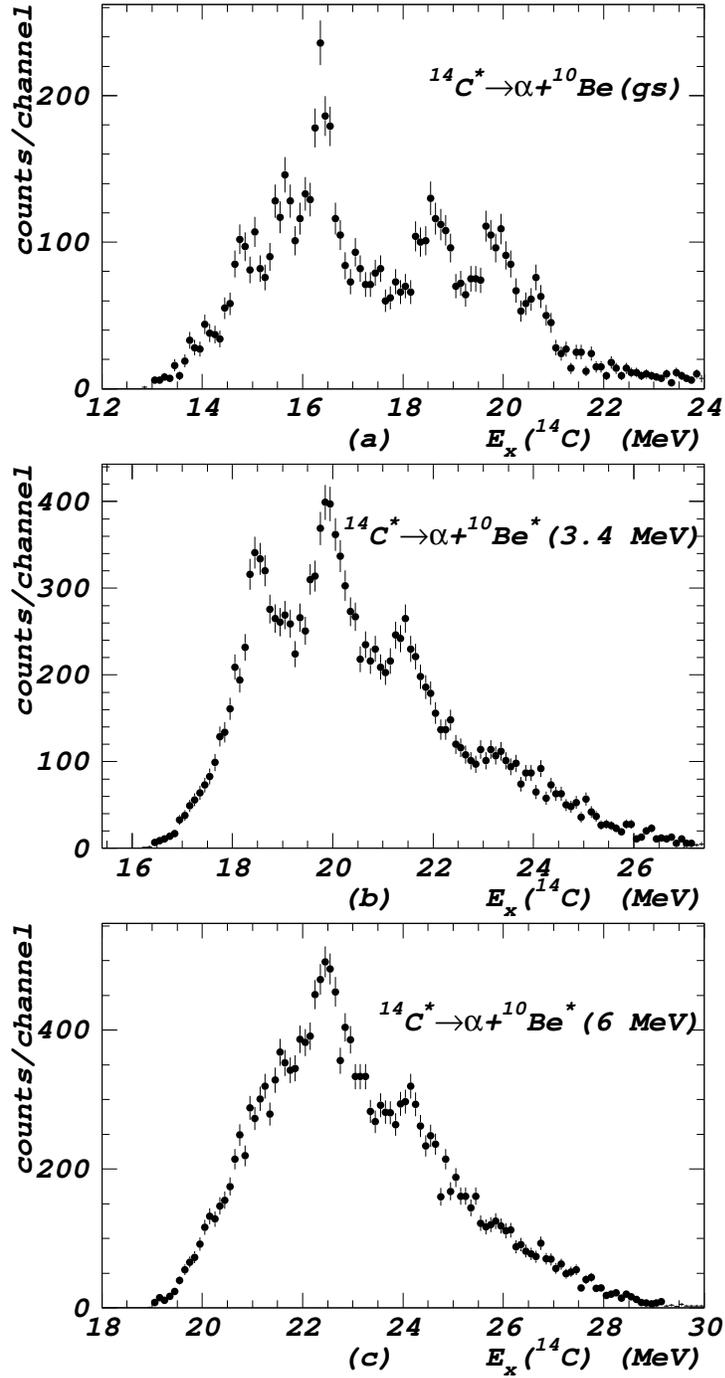}
\caption{\label{exc}$^{14}$C excitation energy spectra for decays to (a) $^{10}$Be
ground state (b) $^{10}$Be 3.4 MeV, 2$^+$ state and (c) $^{10}$Be excited
states at $\sim$6 MeV. Error bars represent statistical errors only.}
\end{figure}

Given the measurement of the momenta of the two fragments, potentially from the
decay of $^{14}$C, it is possible to reconstruct the excitation energy spectrum
of this resonant particle. However, given that there are three particles in the
final state it is possible that they can be produced via the decay of either $^6$Li
into d+$\alpha$ or $^{12}$B into $^{10}$Be+d. Both of these possibilities were
reconstructed and it is clear that there is no, or very little, contribution
from either of these decay processes. Figure \ref{exc}a shows the reconstructed
$^{14}$C excitation energy spectrum for $\alpha$-decays to the $^{10}$Be ground
state, for coincidences between the detectors 1 and 2. The spectrum shows a
series of excited states between 14 and 22 MeV. The spectrum for detectors 3
and 4 covers the excitation energy range 18 to 32 MeV, and shows little
evidence for states beyond 19.8 MeV, indicating that the decay strength is
located at the lower excitation energies. The energies of the peaks are listed
in Table 1. It is possible that some of the peaks in these spectra correspond
to multiple states, however with the present excitation energy resolution
($\sim$300 keV) and statistics it is not possible to be sure if the non-gaussian
peak shapes are due to multiple states or statistical fluctuations.

Figure \ref{exc}b corresponds to $\alpha$-decays to the $^{10}$Be$(2^+)$ state as
measured by the small angle detector pair. Again, the spectrum for the large
angle detectors covers a higher excitation energy region 20 to 34 MeV and
 shows no evidence of additional
structures beyond those observed in Figure \ref{exc}b. Once again, a series of excited
states are observed several of which coincide with those that were observed to
decay to the $^{10}$Be ground state. Finally, the $\alpha$-decay spectrum for
decays to the peak at $\sim$6 MeV in Figure \ref{etot} is shown in Figure \ref{exc}c.
 There is an absolute uncertainty in the excitation energy scale of 300 keV in this
instance given that it is not known which of the four $^{10}$Be excited states
are observed, indeed a combination may be involved. Thus, an excitation energy
of 6 MeV has been assumed for $^{10}$Be in this instance.

Many of the states observed in the present study have possible counterparts in
the present energy level compilation for this nucleus \cite{Ajz91}, as
indicated in Table 1. For example, the states at 14.7, 15.5 and 16.4 MeV
have all been observed in a study of the $^9$Be($^{6}$Li,p)$^{14}$C reaction
\cite{Ajz73}. This reaction is very similar to the present one and is almost
certainly compound in nature, with the evaporation of a proton leading to the
formation of $^{14}$C. In the present case, there is also evidence for a
compound population mechanism as when reconstruction of the $^{14}$C excitation
energy spectrum is performed for the background bump at low total energies in
Figure \ref{etot}, excited states in $^{14}$C may be observed. This may be explained
if the reaction proceeds via the $^{16}$N compound nucleus with the sequential
emission of a proton and neutron to form excited states in $^{14}$C which then
subsequently $\alpha$-decay. Unfortunately, since it is not possible to identify
the excitation energy of the $^{10}$Be for this process it is not possible to
determine the excitation energy of the $^{14}$C states populated via
the n and p sequential emission process. An analysis of the angular distributions
and angular correlations for the states observed in Figure \ref{exc}
using the techniques given in \cite{Fre96} was performed,
but these were found to be essentially featureless, a consequence of the number
of reaction amplitudes contributing to the reaction process due to the presence
of non zero spin particles in the entrance and exit channels. Thus it is not
possible to infer the spins of these states using this reaction.
 The present measurements do however provide the first observation of
$\alpha$-decaying states in this nucleus.

An interesting feature of the data is that the $^{14}$C states which decay to
the ground state and the $^{10}$Be first 2$^+$ state appear to be the same, with the 18.5
and 19.8 MeV states appearing in both spectra. On the other hand, decays to the
6 MeV states in $^{10}$Be appear not to coincide despite there being an overlap
in the excitation energy spectra of Figures \ref{exc}b and \ref{exc}c. There, 
would thus appear to be a preference for the decay of the lower excitation energy states in
$^{14}$C to either the $^{10}$Be ground or first excited state, whilst the
higher lying states (22.4 and 24.0 MeV) decay to the excited states at 6 MeV.
In the absence of spin measurements, it is possible that the decay systematics
reflect changing angular momentum barriers corresponding to decays to states
of differing spins. However, the decay systematics do appear to be more complex
than can be described by such a picture. As stated above, the decays to the
$^{10}$Be ground state and 2$^{+}$ (3.4 MeV) state appear to be similar and thus
are largely unaffected by the change in the decay barrier. The decays to the
2$^{+}$,1$^{-}$,0$^{+}$,2$^{-}$ quartet at $\sim$6MeV, which would sample
similar differences in the angular momenta barriers, highlights different states.
This property, as for example is the case of excited states in $^{20}$Ne
\cite{Hin83,Bro84} and $^{24}$Mg \cite{Led84}, may rather be a reflection of
changing structural overlap of the states in $^{14}$C with those in $^{10}$Be.
This difference may perhaps be understood in terms of the structural content of
the various $^{10}$Be states. In a shell model description the $^{10}$Be ground
state and first excited state correspond to the occupation of the $p_{3/2}$
orbit for the two valence neutrons. On the other hand, the 6 MeV states require
the excitation of one or more of the neutrons to the $s-d$ shell
\cite{von96,von97,von97a}. In the MO \cite{Ita00} and AMD \cite{Kan99}
calculations that have been applied to $^{10}$Be the ground and the first $2^+$ state
correspond to neutrons in $\pi$ like orbits, whereas the 6 MeV states require
either two $\sigma$ neutrons ($0^+_2$) or combinations of $\sigma$ and $\pi$
like neutrons (1$^-$, 2$^-$). Thus, it is possible that the lower energy
$^{14}$C excited states (14.7 to 21.4 MeV) are based upon neutrons in
$\pi$-type molecular configurations, whereas the higher energy states (22.4 and
24.0 MeV) contain some $\sigma$-orbit parentage.
However, it should be noted that there are some states which decay to the 2$^+$
 state that are not strongly observed in decays to the ground state (e.g. 21.4 MeV),
and vice versa (20.6 MeV). This would indicate that other effects may play
 a role in these decay processes, for example angular momentum barriers and
 decay phase space. Whether these states
correspond to single molecular type configurations related by rotational
excitations is unclear, and require measurements of the spins of the states.
Nevertheless, there is a possibility that these states are related to molecular
chains in $^{14}$C.

\section{Summary}
Measurements of the $^7$Li($^{9}$Be,$^4$He$^{10}$Be)$^{2}$H reaction at
$E_{beam}$=70 MeV provide evidence for a series of $^{14}$C excited states
between 14 and 25 MeV. These studies were able to resolve decays to the various
particle bound states in $^{10}$Be. The analysis indicates that decays to the
$^{10}$Be ground state and the first $2^+$ state occur from the same states in $^{14}$C,
whilst a distinct set of states decay to the excited states at $\sim$6 MeV.
Given the proposed molecular content of the states in $^{10}$Be it is speculated
that those in $^{14}$C may also contain molecular characteristics. More definitive
evidence will be provided by future measurements capable of determining the
spins of these excited states.

\begin{acknowledgments}

The authors would like to acknowledge the assistance of ANU personnel in
running the accelerator. This work was carried out under a formal agreement
between the U.K. Engineering and Physical Sciences Research Council and the
Australian National University. PJL, BJG and KLJ would like to acknowledge the
EPSRC for financial support.
\end{acknowledgments}

 \begin{table}
 \caption{\label{14cexc} Excitation energies of $^{14}$C states decaying into states in
$^{10}$Be. The uncertainties in these energies for decays to the $^{10}$Be
ground state and first excited state are 100 keV, and due to the ambiguity in
the excitation energy of the 6 MeV peak, the uncertainty here is 300 keV. The
previous measurements are from the tabulations of \cite{Ajz91}.}
% \begin{ruledtabular}
\begin{tabular}{|c|c|c|c|}\hline
$^{10}$Be$_{gs}$ & $^{10}$Be$(2^+)$ & $^{10}$Be(6 MeV) & Previous\\ \hline
 14.7(0.1) & & & 14.667 (4$^+$) \\ 15.5(0.1) & & & 15.44 (3$^-$) \\ 16.4(0.1) & & & 16.43\\
 18.5(0.1) & 18.5(0.1) & & 18.5\\
  & (19.1(0.1)) & &\\ 19.8(0.1) & 19.8(0.1) & &\\ 20.6(0.1) & & &\\
  & 21.4(0.1) & &\\ & & 22.4(0.3) &\\ & (23.2(0.1)) & &\\ & & 24.0(0.3) &\\\hline
\end{tabular}
%\end{ruledtabular}
\end{table}

\end{document}